\documentclass[twocolumn,aps,prc,superscriptaddress,showpacs,floatfix]{revtex4}
\usepackage{amssymb}
\usepackage{graphicx}

\begin{document}

\title{$D_{sJ}$(2317) meson production at RHIC}
\author{L. W. Chen\footnote{lwchen@sjtu.edu.cn}}
\affiliation{Institute of Theoretical Physics, Shanghai Jiao Tong
University, Shanghai 200240, China}
\author{C. M. Ko\footnote{ko@comp.tamu.edu}}
\affiliation{Cyclotron Institute and Physics Department, Texas
A$\&$M University, College Station, Texas 77843-3366}
\author{W. Liu\footnote{weiliu@comp.tamu.ed}}
\affiliation{Cyclotron Institute and Physics Department, Texas
A$\&$M University, College Station, Texas 77843-3366}
\author{M. Nielsen\footnote{mnielsen@axpfep1.if.usp.br}}
\affiliation{Instituto de F\'{\i}sica, Universidade de S\~ao
Paulo, C.P. 66318, 05315-970 S\~ao Paulo-SP, Brazil}

\begin{abstract}
Production of $D_{sJ}$(2317) mesons in relativistic heavy ion
collisions at RHIC is studied. Using the quark coalescence model, we
first determine the initial number of $D_{sJ}$(2317) mesons produced
during hadronization of created quark-gluon plasma. The predicted
$D_{sJ}$(2317) abundance depends sensitively on the quark structure
of the $D_{sJ}$(2317) meson. An order-of-magnitude larger yield is
obtained for a conventional two-quark than for an exotic four-quark
$D_{sJ}$(2317) meson. To include the hadronic effect on the
$D_{sJ}$(2317) meson yield, we have evaluated the absorption cross
sections of the $D_{sJ}$(2317) meson by pion, rho, anti-kaon, and
vector anti-kaon in a phenomenological hadronic model. Taking into
consideration the absorption and production of $D_{sJ}$(2317) mesons
during the hadronic stage of heavy ion collisions via a kinetic
model, we find that the final yield of $D_{sJ}$(2317) mesons remains
sensitive to its initial number produced from the quark-gluon
plasma, providing thus the possibility of studying the quark
structure of the $D_{sJ}$(2317) meson and its production mechanism
in relativistic heavy ion collisions.
\end{abstract}

\pacs{25.75Nq, 12.39.Fe, 13.75.Lb,14,40.Aq}
\maketitle

\section{Introduction}

A narrow $D_{sJ}$(2317) meson was recently observed by the BABAR
Collaboration \cite{barbar} in the inclusive $D_s^+\pi^0$ invariant
mass distribution from $e^+e^-$ annihilation and confirmed by the
Belle Collaboration in $B$ meson decay \cite{belle}. This meson has
the natural spin-parity $J^P=0^+$ and a mass below what was obtained
from the QCD sum rule approach \cite{hate} and quark model
calculations \cite{godfrey} for a normal two-quark state $c\bar s$.
The $D_{sJ}$(2317) meson has thus been considered as a possible
candidate for the exotic four-quark states that were studied in the
bag model \cite{jaffe,wang}, QCD sum rules \cite{wang1} and the
non-relativistic potential model \cite{weinstein}. It is also
possible that the $D_{sJ}$(2317) meson is simply a $DK$ molecule or
atom.  Determination of the $D_{sJ}$(2317) meson width is limited by
experimental resolutions to a value of less than 4.6 ${\rm MeV}/c^2$
\cite{belle}. The small width of $D_{sJ}$(2317) meson is not
surprising as its mass is below the threshold of $DK$ system and can
only decay into the kinematically allowed but isospin violated
channel of $D_s\pi$ state. Theoretically, the decay width of
$D_{sJ}\to D_s\pi$ has been studied using the QCD sum rules, and its
value varies from a few keV \cite{marina1} to a few tens keV
\cite{weiwei} depending on the assumed flavor state of four quarks
or two quarks, respectively. A more phenomenological approach based
on the $^3P_0$ model \cite{jielu} also gives a narrow width of about
a few tens keV for a normal two-quark $D_{sJ}$(2317) meson.

Studying the mechanism for $D_{sJ}$(2317) meson production in
nuclear reactions is useful for understanding its quark structure.
In Ref.\cite{george}, a coupled-channel quark model has been used to
study the production of a two-quark $D_{sJ}$(2317) meson and its
radial excitations in hadronic reactions. Production of
$D_{sJ}$(2317) in relativistic heavy ion collisions has also been
studied \cite{zhongbiao}. It was found that for a four-quark
$D_{sJ}$(2317) meson, a much large yield is obtained if one takes
into account the diquark-diquark interactions in the produced
quark-gluon plasma. Since the $D_{sJ}$(2317) meson is not expected
to survive in the quark-gluon plasma, it is more likely to be
produced at hadronization of the quark-gluon plasma either
statistically or via quark coalescence. Its final abundance in a
heavy ion collision depends, however, also on its absorption and
production probability in the subsequent hadronic matter.

In the present paper, we study $D_{sJ}$(2317) meson production in
central heavy ion collisions at the Relativistic Heavy Ion Collider
(RHIC) in a kinetic model that starts from the final stage of the
quark-gluon plasma, goes through a mixed phase of quark-gluon and
hadronic matters, and finally undergoes the hadronic expansion. The
production of $D_{sJ}$(2317) mesons from the quark-gluon plasma is
modeled by the constituent quark coalescence model, which has been
shown to describe reasonably not only the particle yields and their
ratios \cite{alcor} but also their transverse momentum spectra and
anisotropic flows \cite{greco,hwa,fries}. The predicted number of
$D_{sJ}$(2317) mesons is found to depend on its quark structure,
with the two-quark state giving an order-of-magnitude larger value
than the four-quark state. The $D_{sJ}$(2317) meson can be absorbed
and also produced in subsequent hadronic matter via the reactions
$\pi D_{sJ}\leftrightarrow K^*D(KD^*)$, $\rho D_{sJ}\leftrightarrow
KD$, $\bar K D_{sJ}\leftrightarrow \rho D(\pi D^*)$, and $\bar K^*
D_{sJ}\leftrightarrow\pi D$. The cross sections for these reactions
are evaluated in a phenomenological hadronic model with coupling
constants and form factors involving the $D_{sJ}$(2317) meson
determined from the QCD sum rules. Taking into account the hadronic
effect in heavy ion collisions via a kinetic approach, we find that
the final number of $D_{sJ}$(2317) mesons remains sensitive to its
initial number produced from the quark-gluon plasma. Studying
$D_{sJ}$(2317) meson production in heavy ion collisions thus
provides the possibility to study both its production mechanism and
quark structure.

This paper is organized as follows. In Section \ref{dynamics}, the
dynamics of heavy ion collisions at RHIC is described. Production of
$D_{sJ}$(2317) meson from the initial quark-gluon plasma via the
quark coalescence model is discussed in Section \ref{coalescence}.
The absorption cross sections of the $D_{sJ}$(2317) meson by hadrons
such as the pion, rho, anti-kaon, and vector anti-kaon as well as
their thermally averaged values are evaluated in Section
\ref{hadronic}. Solving the rate equation based on a kinetic model,
time evolution of the $D_{sJ}$ (2317) meson abundance in heavy ion
collisions is presented in Section \ref{kinetic}. A summary is then
given in Section \ref{summary}. Details on the derivation of an
approximate analytical coalescence formula for $D_{sJ}$(2317) meson
production from the quark-gluon plasma is given in Appendix
\ref{analytic}, while those on the QCD sum-rule approach to the
determination of the form factor at the $D_{sJ}DK$ vertex, which is
needed in calculating the $D_{sJ}$(2317) meson absorption cross
sections, is described in Appendix \ref{qcd}.

\section{Heavy ion collision dynamics at RHIC}
\label{dynamics}

To model the dynamics of central relativistic heavy ion collisions
after the end of the quark-gluon plasma phase, we use the schematic
model of Ref.\cite{liewen} based on the boost invariant picture of
Bjorken \cite{bjorken} and an accelerated transverse expansion.
Specifically, the volume of produced fire-cylinder in central Au+Au
collisions at $\sqrt{s_{NN}}=200$ GeV, which is the collision we are
interested in, is taken to evolve with the proper time according to
\begin{equation}
V(\tau )=\pi \left[ R_{\mathrm{C}}+v_{\mathrm{C}}(\tau -\tau
_{\mathrm{C}})+a/2(\tau -\tau _{\mathrm{C}})^{2}\right] ^{2}\tau c,
\end{equation}%
where $R_{\mathrm{C}}=8~{\rm fm}$ and $\tau _{\mathrm{C}}=5~{\rm
fm}/c$ are final transverse and longitudinal sizes of the
quark-gluon plasma, while $v_{\rm C}=0.4c$ is its transverse flow
velocity at this time. The total transverse energy of quarks and
gluons in the midrapidity ($|y|\leq 0.5$) is then about $1,067$ GeV
if quarks and gluons are taken to be massive with $m_{g}=500$ MeV,
$m_{u}=m_{d}\equiv m_q=300$ MeV and $m_{s}=475$ MeV in order to take
into account the nonperturbative effects of QCD near the critical
temperature \cite{levai}, and if the quark strangeness and baryon
chemical potentials are taken to be $\mu _{s}=0$ and $\mu _{b}=10$
MeV, respectively, to account for strangeness neutrality in the
quark-gluon plasma and observed final antiproton to proton ratio of
about 0.7 at RHIC. The fire-cylinder then goes through a mixed phase
of partonic and hadronic matters at a constant temperature $T_{\rm
C}$ until $\tau_H=7.5$ fm/$c$ when its transverse radius and flow
velocity are $R_H\approx 9$ fm and $v_H\approx 0.45c$, respectively.
This corresponds to a small transverse acceleration $a=0.02$
$c^{2}$/fm, which is chosen to reflect the small pressure near phase
transition and in hadronic matter \cite{karsh} as well as to obtain
a lifetime for the expanding matter comparable to that from the
transport model \cite{ampt}. Values of other parameters of the
fire-cylinder are determined from fitting the measured transverse
energy $\simeq 788$ GeV as well as the extracted freeze out
temperature $T_F=125$ MeV and transverse flow velocity $\simeq
0.65c$ of midrapidity hadrons in central Au+Au collisions at
$\sqrt{s_{NN}}=200$ GeV. Assuming that the fire-cylinder expands
isentropically, the fraction of hadronic matter during the mixed
phase is found to increase approximately linearly, and the time
dependence of the temperature of the fire-cylinder obtained in
Ref.\cite{liewen} can be parameterized as
\begin{eqnarray}\label{temperature}
T(\tau)=T_C-(T_H-T_F)\left(\frac{\tau-\tau_H}{\tau_F-\tau_H}\right)^{0.8},
\end{eqnarray}
where $T_H$ is the temperature of the hadronic matter at the end of
the mixed phase and is thus the same as the critical temperature
$T_C$ for the quark-gluon plasma to hadronic matter transition. As
in Ref.\cite{liewen}, we take $T_H=T_C=175$ MeV. The freeze out
temperature $T_F=125$ MeV then leads to a freeze out time
$\tau_F\approx 17.3$ fm/$c$.

\begin{figure}[th]
\includegraphics[height=3.2in,width=2.8in]{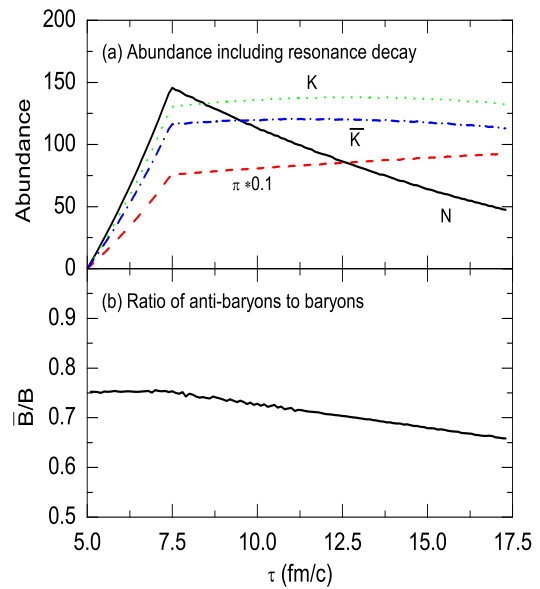}
\caption{{\protect\small (Color online) Time evolution of the
abundance of pions, kaons, anti-kaons, and nucleons including
contributions from decays of resonances (a) and the ratio of
anti-baryon to baryon abundances (b) of mid-rapidity particles in
central Au+Au collisions at
}$\protect\sqrt{s_{NN}}=200${\protect\small \ GeV.}} \label{PiKN}
\end{figure}

For normal hadrons such as pions, kaons, anti-kaons, and nucleons,
they are taken to be in chemical equilibrium with baryon chemical
potential $\mu _{B}=30$ \textrm{MeV}, charge chemical potential $\mu
_{Q}=0$ \textrm{MeV}, and strangeness chemical potential $\mu
_{S}=10$ \textrm{MeV}. The nonzero strange chemical potential is
needed to account for observed $K^-/K^+\approx 0.9$ ratio in heavy
ion collisions at RHIC. Neglecting the time dependence of the
chemical potentials, which have been shown to vary weakly with the
temperature of an isentropically expanding matter in heavy ion
collisions at RHIC \cite{rapp}, time evolution of the abundance of
pions, kaons, anti-kaons, and nucleons has been shown in Fig.~1 of
Ref.\cite{liewen}. Including also contributions from decays of
resonances, time evolution of the abundance of these hadrons is
shown in Fig.\ref{PiKN}(a).  It is seen that the total numbers of
pions, kaons and anti-kaons do not change much during the hadronic
stage while the nucleon number decreases significantly as
temperature drops. Their final numbers at freeze out are $926$,
$133$, $113$, and $47$, respectively, and are comparable to those
measured in experiments. In Fig.\ref{PiKN}(b), the time evolution of
the ratio of anti-baryons to baryons is shown, and it changes from
an initial value of $0.75$ to a final value of $0.66$, which is also
comparable to the measured value of about $0.7$.

\section{$D_{sJ}$(2317) meson production from the quark-gluon plasma}
\label{coalescence}

\subsection{The coalescence model}

In the coalescence model, the number of $D_{sJ}$(2317) mesons that
are produced from the quark-gluon plasma is given by the product of
a statistical factor $g_{D_{sJ}}$ which denotes the probability of
combining $c\bar{s}$ or $c\bar{s}q\bar{q}$ quarks into a color
neutral, spin 0, and isospin 0 hadronic state and depends on whether
the $D_{sJ}$(2317) meson is a two-quark or four-quark state, and the
overlap of the quark phase-space distribution function
$f_{q}(x_{i},p_{i})$ in the fire-cylinder with the Wigner
distribution function $f_{D_{sJ}}^W$ of the $D_{sJ}$(2317) meson.
The latter corresponds to the probability of converting the above
partonic state into $D_{sJ}$(2317) and is dependent of the quark
spatial wave functions in the $D_{sJ}$(2317) meson. Explicitly, the
$D_{sJ}$(2317) number is expressed as
\begin{eqnarray}\label{coalmodel}
N_{D_{sJ}}^{\rm coal} &=&g_{D_{sJ}}\int_{\sigma_C}
\prod_{i=1}^{n}\frac{p_{i}\cdot d\sigma
_{i}d^{3}\mathbf{p}_{i}}{(2\pi )^{3}E_{i}}f_{q}(x_{i},p_{i})
\nonumber \\
&&\times f_{D_{sJ}}^W(x_{1}..x_{n};p_{1}..p_{n}),  \label{coal}
\end{eqnarray}
with $n=2$ or $4$ for a two-quark or four-quark $D_{sJ}$(2317).
Similar expressions have previously been used for studying the
production of strange hadrons \cite{strange}, charmed mesons
\cite{charm}, and penta-quark baryons \cite{liewen} from the
quark-gluon plasma formed in relativistic heavy-ion collisions.

We note that the coalescence model can be viewed as formation of
bound states from interacting particles with energy mismatch
balanced by other particles in the system. Neglecting such off-shell
effects is reasonable if the binding energy is not large and/or if
the production process is fast compared to the inverse of the energy
mismatch.

Since the normalized quark wave function of the $D_{sJ}$(2317) meson
in the color-spin-isospin space can be expressed as a linear
combination of all possible orthogonal flavor, color, and spin basis
states, with coefficients depending on the quark model used for the
$D_{sJ}$(2317) meson, $g_{D_{sJ}}$ is simply given by the
probability of finding these quarks in any one of these
color-spin-isospin basis states, i.e., $g_{D_{sJ}}=1/3^2\times
1/2^2=1/36$ or $1/3^{4}\times 1/2^{4}=1/1296$ for a two-quark or
four-quark $D_{sJ}$(2317) meson, respectively.

The $d\sigma$ in Eq.(\ref{coal}) denotes an element of a space-like
hypersurface $\sigma_C$ at hadronization \cite{cooper}. In terms of
the proper time $\tau=(t^2-z^2)^{1/2} $, the longitudinal
momentum-energy rapidity $y=\frac{1}{2}\ln(\frac{E+p_z}{E-p_z})$ and
space-time rapidity $\eta=\frac{1}{2}\ln(\frac{t+z}{t-z})$, the
polar coordinates $r$ and $\phi$ in the transverse plane, the
covariant volume element can be written as $p\cdot d\sigma=\tau
m_{T}\cosh (y-\eta )rdrd\phi d\eta$.

For the phase-space distribution functions of quarks in the
fire-cylinder, they are taken to be the same as in the description
of the heavy ion collision dynamics, i.e., they are uniformly
distributed in the transverse plane and their momentum distributions
are relativistic Boltzmannian in the transverse direction but
uniform in rapidity along the longitudinal direction. Also, the
Bjorken correlation of equal spatial $\eta$ and momentum $y$
rapidities is imposed, which is consistent with the small difference
$|y-\eta|\le 0.5$ seen in the transport model \cite{ampt}.
Explicitly, the quark momentum distribution per unit rapidity at
$T_C$ is
\begin{eqnarray}\label{quark}
f_q(\eta, {\bf r},y,\mathbf{p}_T)&=&g_q\delta(\eta-y)\nonumber\\
&\times&\exp\left(\frac{-\gamma(m_T-{\bf p}_T\cdot{\bf \beta
})-\mu_q}{T_C}\right),
\end{eqnarray}
where $g_q=6$ is the color-spin degeneracy of a quark, ${\bf
\beta}=({\bf r}/R)v_C$ is the radial-dependent transverse flow
velocity, $\gamma=(1-\beta^2)^{-1/2}$, and the quark chemical
potential $\mu_q=\mu_b+\mu_s$. The abundances of quarks at the end
of the QGP phase in central Au+Au collisions at $\sqrt{s_{NN}}=200$
GeV described by the expanding fire-cylinder model of previous
section are $N_{u}=N_{d}\simeq 245$, and $N_{\bar{s}}\simeq 149$ at
$\tau _{\mathrm{C}}$, if we take into account the effect of gluons
by converting them into quarks according to the quark flavor
composition in the quark-gluon plasma as in Ref.\cite{greco}. For
charm quarks, we assume that they are in thermal equilibrium in the
quark-gluon plasma, which is supported by the large elliptic flow of
electrons from charmed meson decays that are observed in experiments
\cite{Adler:2005ab,Laue:2004tf} and transport models
\cite{bzhang,molnar}. Their number is, however, at present uncertain
as it depends on whether charm quarks can be produced from the
quark-gluon plasma. If we assume that the latter contribution is
unimportant, then the number of charm quarks $N_c$ produced in heavy
ion collisions is simply given by the product of the charm quark
number $N_c^{NN}$ produced from an initial hard nucleon-nucleon
scattering and the total number of binary collisions (about 960) in
central Au+Au collisions. Using $N_c^{NN}$ from the PYTHIA program,
we obtain $N_c\sim 1.5$. The value of $N_c$ increases to about 3 and
7 if we use the $N_C^{NN}$ from the PHENIX \cite{phenix} and the
STAR experiment \cite{star}, respectively. In the following study,
we will use $N_c=3$ for the calculation and discuss the sensitivity
of our results to the change in the value for $N_c$. For the charm
quark mass, it is taken to be $m_c=1.5$ GeV.

For the Wigner distribution function of the $D_{sJ}$(2317) meson,
instead of using a function of Lorentz invariant four-dimensional
relative coordinates and momenta such as in Refs.\cite{dover,greco},
we take it to be a function of three-dimensional relative
coordinates and momenta for simplicity, based on the harmonic
oscillator wave functions. Specifically, it is obtained from
assuming that the wave functions of the quarks are those of a
harmonic oscillator with an oscillator frequency $\omega$. For a
two-quark $D_{sJ}$(2317) meson with $J^\pi=0^+$, its $c\bar s$
quarks are in the relative $p$-wave and its Wigner distribution
function is thus \cite{dover}
\begin{eqnarray}\label{wigner1}
f_{D_{sJ}}^{W,{\rm two}}(x;p)&=&\left(\frac{16}{3}\frac{{\bf
y}^2}{\sigma^2}-8+\frac{16}{3}\sigma^2{\bf k}^2\right)\nonumber\\
&\times&\exp\left(-\frac{{\bf y}^2}{\sigma^2}-\sigma^2{\bf
k}^2\right).
\end{eqnarray}
In the above, we have used the usual definitions ${\bf y}={\bf
x_1}-{\bf x_2}$ and ${\bf k}=(m_s{\bf p_1}-m_c{\bf p}_2)/(m_c+m_s)$
for the relative coordinate and momentum of the two quarks,
respectively. For the width parameter $\sigma$, it can be related to
the oscillator frequency $\omega$ by $\sigma=1/(\mu\omega)^{1/2}$
with the reduced mass $\mu$ given by $\mu=m_cm_s/(m_c+m_s)$.

If the $D_{sJ}$(2317) meson is a four-quark meson, its four quarks
are all in relative s-waves, which then leads to the Wigner
distribution function
\begin{equation}\label{wigner2}
f_{D_{sJ}}^{W,{\rm four}}(x;p)=8^3\exp \left(-\sum_{i=1}^{3}
\frac{\mathbf{y}_{i}^{2}}{\sigma_{i}^{2}}
-\sum_{i=1}^{3}\mathbf{k}_{i}^{2}\sigma_{i}^{2}\right),
\end{equation}
where the relative coordinates $\mathbf{y}_{i}$ and momenta
$\mathbf{k}_{i}$ are related to the quark coordinates
$\mathbf{x}_{i}$ and momenta $\mathbf{p}_{i}$ by the Jacobian
transformations:
\begin{eqnarray}\label{relative1}
{\bf y}_1&=&\frac{{\bf x}_1-{\bf x}_2}{\sqrt{2}},\nonumber\\
{\bf y}_2&=&\sqrt{\frac{2}{3}}\left(\frac{m_c}{m_c+m_s}{\bf
x}_1+\frac{m_s}{m_c+m_s}{\bf x}_2-{\bf x}_3\right),\nonumber\\
{\bf y}_3&=&\sqrt{\frac{3}{4}}\left(\frac{m_c}{m_c+m_s+m_q}{\bf x}_1
+\frac{m_s}{m_c+m_s+m_q}{\bf x}_2\right.\nonumber\\
&+&\left.\frac{m_q}{m_c+m_s+m_q}{\bf x}_3-{\bf x}_4\right),
\end{eqnarray}
and
\begin{eqnarray}\label{relative2}
{\bf k}_1&=&\sqrt{2}\frac{m_s{\bf p}_1-m_c{\bf p}_2}{m_c+m_s},\nonumber\\
{\bf k}_2&=&\sqrt{\frac{3}{2}}\frac{m_q({\bf p}_1+{\bf
p}_2)-(m_c+m_s){\bf p}_3}{m_c+m_s+m_q},\nonumber\\
{\bf k}_3&=&\sqrt{\frac{4}{3}}\frac{m_q({\bf p}_1+{\bf p}_2+{\bf
p}_3)-(m_c+m_s+m_q){\bf p}_4}{m_c+m_s+2m_q}.
\end{eqnarray}
It can be shown that the product of the Jacobians for the coordinate
and momentum transformations is equal to one.

The width parameter $\sigma_{i}$ for the $i$-th relative coordinate
in a four-quark $D_{sJ}$(2317) meson is again given by $\sigma
_{i}=1/(\mu _{i}\omega) ^{1/2}$ with the reduced masses
\begin{eqnarray}
\mu_1&=&\frac{2m_cm_s}{m_c+m_s},\nonumber\\
\mu_2&=&\frac{3}{2}\frac{m_q(m_c+m_s)}{m_c+m_s+m_q},\nonumber\\
\mu_3&=&\frac{4}{3}\frac{m_q(m_c+m_s+m_q)}{m_c+m_s+2m_q}.
\label{reducedmass}
\end{eqnarray}
We note that the reduced mass $\mu_1$ of the $c\bar s$ quark pair in
a four-quark $D_{sJ}$(2317) meson is a factor of two larger than
that in a two-quark $D_{sJ}$(2317) meson due to differences in the
definitions of the relative coordinate and momentum.

\subsection{Number of $D_{sJ}$(2317) mesons produced from the quark-gluon plasma}

To evaluate the number of $D_{sJ}$(2317) mesons produced from the
quark-gluon plasma requires information on the oscillator frequency
$\omega$ through the width parameter $\sigma$ in the $D_{sJ}$(2317)
meson Wigner distribution function, which is related to the size of
$D_{sJ}$(2317). Since the latter is not known empirically, we choose
the value of the oscillator frequency to fit instead the
root-mean-square charge radius of the s-wave charmed $D_s^+(c\bar
s)$ meson. Taking its Wigner distribution function similar to
Eq.(\ref{wigner2}) but with only one relative coordinate and
momentum, we obtain the following mean-squared charge radius for the
$D_s^+(c\bar s)$ meson:
\begin{eqnarray}
\langle r_{D_s}^2\rangle_{\rm ch}&=&\frac{2}{3}\langle({\bf
x}_1-{\bf Y})^2\rangle+\frac{1}{3}\langle({\bf x}_2-{\bf Y})^2\rangle\nonumber\\
&=&\frac{m_c^2+2m_s^2}{3(m_c+m_s)^2}\langle {\bf y}^2\rangle=
\frac{m_c^2+2m_s^2}{2(m_c+m_s)^2}\sigma^2.
\end{eqnarray}
In the above, ${\bf Y}=(m_c{\bf x}_1+m_s{\bf x}_2)/(m_c+m_s)$ is the
center-of-mass coordinate of $c\bar s$ quarks, and we have used the
relation $\langle {\bf y}^2\rangle=(3/2)\sigma^2$ between the
mean-square distance and the width parameter for two quarks in the
relative s-wave as in the $D_s^+(c\bar s)$ meson. Using the value
$\langle r_{D_s}^2\rangle_{\rm ch}\approx 0.124~{\rm fm}^2$
determined from the light-front quark model \cite{hwang}, we find
that $\sigma\approx 0.60~{\rm fm}$ and $\hbar\omega\approx 300~{\rm
MeV}$.

For a two-quark $D_{sJ}$(2317) meson, whose quarks are in the
relative p-wave, the relation between the mean-square distance of
the two quarks and the width parameter is $\langle{\bf
y}^2\rangle=(5/2)\sigma^2$. Using above determined width parameter
$\sigma$, we obtain the following root-mean-square radius for a
two-quark $D_{sJ}$(2317) meson:
\begin{eqnarray}
\langle r_{D_{sJ}}^2\rangle^{1/2}_{\rm two}&=&\frac{1}{\sqrt{2}}
\frac{(m_c^2+m_s^2)^{1/2}}{m_c+m_s}\langle{\bf y}^2\rangle^{1/2}\nonumber\\
&=&\frac{\sqrt{5}}{2}\frac{(m_c^2+m_s^2)^{1/2}}{m_c+m_s}\sigma
\approx 0.53~{\rm fm}.
\end{eqnarray}

For a four-quark $D_{sJ}$(2317) meson, its three size parameters are
$\sigma_1=1/(\mu_1\omega)^{1/2}\approx 0.42$ fm,
$\sigma_2=1/(\mu_2\omega)^{1/2}\approx 0.58$ fm, and
$\sigma_3=1/(\mu_3\omega)^{1/2}\approx 0.6$ fm. The resulting
root-mean-square radius of a four-quark $D_{sJ}$(2317) meson is then
\begin{eqnarray}
\langle r_{D_{sJ}}^{2}\rangle^{1/2}_{\rm four}&=&
\left[\frac{3}{4}\frac{(m_c^2+m_s^2)\sigma_1^2}{(m_c+m_s)^2}\right.\nonumber\\
&+&\frac{9}{16}\frac{((m_c+m_s)^2+2m_q^2)\sigma_2^2}{(m_c+m_s+m_q)^2}\nonumber\\
&+&\frac{1}{2}\left.\frac{((m_c+m_s+m_q)^2+3m_q^2)\sigma_3^2}
{(m_c+m_s+2m_q)^2}\right]^{1/2}\nonumber\\
&\approx& 0.62~\mathrm{fm},
\end{eqnarray}
which is somewhat larger than that of a two-quark $D_{sJ}$(2317)
meson.

The coalescence integral in Eq.(\ref{coalmodel}) can be evaluated
analytically if we expand the hyperbolic functions to first order,
neglect the transverse flow, and use non-relativistic momentum
distributions for quarks. The first approximation is valid for
$|y|\leq 0.5$ considered in present study. Although the transverse
flow affects strongly the transverse momentum spectrum of produced
$D_{sJ}$(2317) mesons, it only has a small effect on its number. As
shown in Appendix \ref{analytic}, these approximations lead to the
following numbers of produced $D_{sJ}$(2317) mesons from quark
coalescence: $\sim 1.9\times 10^{-2}$ for a two-quark $D_{sJ}$(2317)
meson and $\sim 1.1\times 10^{-3}$ for a four-quark $D_{sJ}$(2317)
meson. These numbers are about a factor of two larger than those
obtained from numerically evaluating the coalescence integral using
the Monte Carlo method of Ref.\cite{greco}, which gives about
$9.8\times 10^{-3}$ and $4.2\times 10^{-4}$ per unit rapidity for
the two-quark and four-quark $D_{sJ}$(2317) mesons, respectively,
largely due to the use of the relativistic quark distribution
functions.

Since equilibrium thermal models have been successfully employed in
describing the experimental data for the yields and ratios of many
hadrons in heavy ion collisions at RHIC \cite{braun,rafelski}, it is
of interest to compare the predicted number of $D_{sJ}$(2317) mesons
from the coalescence model with that from the statistical model. In
terms of the charm fugacity $\gamma_C$ and the strangeness chemical
potential $\mu_S$, this model gives the following number of produced
$D_{sJ}$(2317) mesons at hadronization:
\begin{eqnarray}\label{statistical}
N_{D_{sJ}}^{\rm stat}&=\gamma_C&\int_{\sigma
_{h}}\frac{p^{\mu }d\sigma_{\mu }}{(2\pi)^{3}}\frac{d^{3}{\bf
p}}{E}f_{D_{sJ}}(x,p)\nonumber \\
&\approx&\frac{V_H\gamma_C e^{\mu_S/T_H}}{(2\pi)^2}\int dm_T
m_{T}^{2}e^{-\frac{{\bar\gamma}_H m_T}{T_H}}\nonumber\\
&\times&I_{0}\left(\frac{{\bar\gamma}_H{\bar\beta}_H
p_{T}}{T_C}\right)\approx 5.2\times 10^{-2},
\end{eqnarray}
where $f_{D_{sJ}}(x,p)$ is the thermal distribution function of
$D_{sJ}$(2317) mesons, given by an expression similar to
Eq.(\ref{quark}) for quarks and $I_0$ is the modified Bessel
function. In obtaining the numerical value in the last line of above
equation, we have used $V_H\approx 1,908~{\rm fm}^3$, $T_H=175~{\rm
MeV}$, ${\bar\beta}_H=0.3c$, $\mu_S=10~{\rm MeV}$, and the charm
fugacity $\gamma_C\approx 8.4$. The latter ensures that the numbers
of charmed hadrons produced statistically at hadronization is same
as the number of charm quarks $N_c$ in the quark-gluon plasma.
Specifically, we have $N_D\approx 1.1$, $N_{D^*}\approx 1.5$,
$N_{D_s}\approx 0.31$, and $N_{\Lambda_c}\approx 0.11$, giving a
total of about 3 charmed hadrons.  We note that the number of
$D_{sJ}$(2317) mesons produced in the statistical model is
independent of its quark structure, contrary to that in the
coalescence model in which the yield for a two-quark $D_{sJ}$(2317)
meson is about a factor of twenty larger than that for a four-quark
one.

\section{Hadronic effects on the $D_{sJ}$(2317) meson}

\label{hadronic}

\subsection{$D_{sJ}$(2317) meson absorption cross sections by hadrons}

\begin{figure}[ht]
\includegraphics[width=3in,height=2.5in,angle=0]{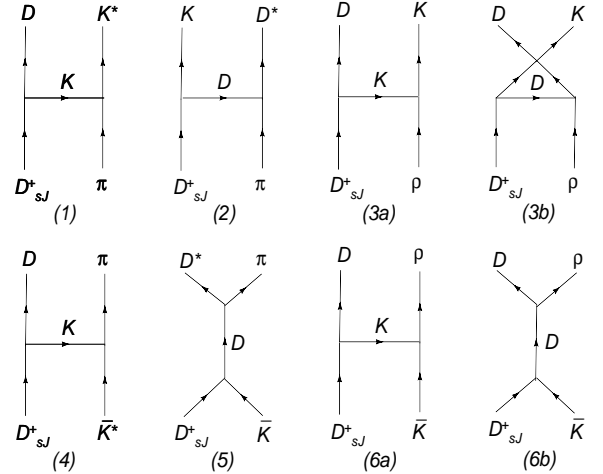}
\caption{Born diagrams for $D_{sJ}$(2317) absorption by
$\protect\pi$, $\protect\rho$, ${\bar K}$, and ${\bar K}^*$ mesons.}
\label{diagram}
\end{figure}

The abundance of $D_{sJ}$(2317) mesons can change during the
expansion of the hadronic matter as a result of absorption by pions,
rho mesons, anti-kaons, and vector anti-kaons. Neglecting reactions
with a $D_s$ meson in the final states, which are suppressed as a
result of the presence of the isospin violated vertex
$D_{sJ}D_s\pi$, we have the following reactions:
\begin{eqnarray}
&&\pi D_{sJ}\to KD^*(K^*D),~~ \rho D_{sJ}\to KD,  \nonumber \\
&&\bar KD_{sJ}\to\rho D(\pi D^*),~~ \bar K^* D_{sJ}\to \pi D,
\end{eqnarray}
as shown in Fig.\ref{diagram} for the lowest-order Born diagrams.
The cross sections for these reactions can be evaluated using the
interaction Lagrangians
\begin{eqnarray}  \label{intlag}
\mathcal{L}_{\rho KK}&=& ig_{\rho KK} (\bar{K}\vec{\tau}\partial_\mu K-
\partial_\mu \bar{K}\vec{\tau}K) \cdot\vec{\rho}^\mu,  \nonumber \\
\mathcal{L}_{\rho DD}&=& -ig_{\rho DD}
(\bar{D}\vec{\tau}\partial_\mu D-
\partial_\mu \bar{D}\vec{\tau}D) \cdot\vec{\rho}^\mu,  \nonumber \\
\mathcal{L}_{K^*K\pi}&=&ig_{K^* K\pi}\bar{K}_\mu^* \vec{\tau}%
\cdot(K\partial^\mu \vec{\pi}-\partial^\mu K\vec{\pi}) \mathrm{+H.c.},
\nonumber \\
\mathcal{L}_{D^*D\pi}&=&-ig_{D^* D\pi}\bar{D}_\mu^* \vec{\tau}%
\cdot(D\partial^\mu \vec{\pi}-\partial^\mu D\vec{\pi}) \mathrm{+H.c.},
\nonumber \\
\mathcal{L}_{D_{sJ}DK}&=& g_{D_{sJ}DK}KDD_{sJ}.
\end{eqnarray}
In the above, $\vec{\tau}$ are Pauli matrices for isospin; and
$\vec{\pi}$ and $\vec{\rho}$ denote the pion and rho meson isospin
triplet, respectively; and $K=(K^+ ,K^0)^T$ and $K^* =(K^{*+}
,K^{*0})^T$ denote the pseudoscalar and vector strange meson isospin
doublet, respectively. The isospin doublet pseudoscalar $D$ and
vector $D^*$ mesons are defined in a similar way. For coupling
constants, we use $g_{\rho DD}=2.52$ from the vector dominance model
(VDM) \cite{linko,linkozhang}, $g_{K^*K\pi}=3.25$ \cite{koseibert}
and $g_{D^*D\pi}=6.3$ \cite{cleo} from the decay widths of $K^*$ and
$D^*$, respectively, and $g_{\rho KK}=3.25$ from the $SU(3)$
symmetry \cite{chli}. For the coupling constant $g_{KDD_{sJ}}$, it
has been studied in the QCD sum rules, and its value depends
strongly on the quark structure of $D_{sJ}$(2317) meson. The
predicted values are $g_{D_{sJ}DK}=9.2$ GeV if the $D_{sJ}$(2317)
meson is a two-quark state \cite{wang2} and $g_{D_{sJ}DK}=3.15$ GeV
if it is a four-quark state \cite{marina2}.

The amplitudes for the reactions shown in Fig.\ref{diagram} are
given by
\begin{eqnarray}
\mathcal{M}_{1}&=&-\tau^a_{ij}g_{D_{sJ}DK}g_{K^*K\pi}\frac{1}{t-m^2_K}
(2p_2-p_4)_{\mu}\epsilon^\mu_4,  \nonumber \\
\mathcal{M}_{2}&=&\tau^a_{ij}g_{D_{sJ}DK}g_{D^*D\pi}\frac{1}{t-m^2_D}
(2p_2-p_4)_{\mu}\epsilon^\mu_4,  \nonumber \\
\mathcal{M}_{3a}&=&\tau^a_{ij}g_{D_{sJ}DK}g_{\rho KK}
\frac{1}{t-m^2_K}(2p_4-p_2)_{\mu}\epsilon^\mu_2,  \nonumber \\
\mathcal{M}_{3b}&=&\tau^a_{ij}g_{D_{sJ}DK}g_{\rho DD}
\frac{1}{u-m^2_D}(p_2-2p_3)_{\mu}\epsilon^\mu_2,  \nonumber \\
\mathcal{M}_{4}&=&\mathcal{M}_{1}(p_2\leftrightarrow -p_4),  \nonumber \\
\mathcal{M}_{5}&=&\mathcal{M}_{2}(p_2\leftrightarrow
-p_3;p_3\leftrightarrow p_4),  \nonumber \\
\mathcal{M}_{6}&=&\mathcal{M}_{3}(p_2\leftrightarrow -p_4).
\end{eqnarray}
In the above, the matrix element $\tau^a_{ij}$ takes into account
the isospin states of the particles in a reaction, with $a$ denoting
those of isospin triplet $\pi$ and $\rho$ mesons, and $i$ and $j$
those of isospin doublet $K$, $K^*$, $D$, and $D^*$ mesons. The
momenta $p_{1}$ and $p_{2}$ are those of initial state particles
while $p_{3}$ and $p_{4}$ are those of final state particles on the
left and right side of a diagram. The usual Mandelstam variables are
given by $s=(p_{1}+p_{2})^{2}$, $t=(p_{1}-p_{3})^{2} $, and
$u=(p_{1}-p_{4})^{2}$.

To obtain the full amplitudes, one needs in principle to carry out a
coupled-channel calculation in order to avoid violation of
unitarity. Such an approach is, however, beyond the scope of present
study. To prevent the artificial growth of the tree-level amplitudes
with the energy, we introduce instead form factors at interaction
vertices, which are taken to have the form \cite{weiliu}
\begin{eqnarray}  \label{form1}
F(\mathbf{q}) =\frac{\Lambda^2}{\Lambda^2 +\mathbf{q}^2},
\end{eqnarray}
where $\mathbf{q}^2$, taken in the center of mass, is the squared
three momentum transfer for $t$ and $u$ channels, or the squared
three momentum of either the incoming or outgoing particles for $s$
channel. For the cutoff parameter $\Lambda$, we use $\Lambda=1.3$
GeV for vertices involving an off-shell $K$ meson and $\Lambda=3.7$
GeV for those involving an off-shell $D$ meson. These values are
determined from the QCD sum-rule calculations given in Appendix
\ref{qcd} for the $D_{sJ}DK$ three-point functions. Although the
calculations are only for a two-quark $D_{sJ}(2317)$ meson, we use
them also for a four-quark $D_{sJ}(2317)$ meson as well as for other
vertices in the diagrams in Fig.~\ref{diagram}. We expect this to be
a reasonable assumption as a study of the $X(3872)$ meson in the QCD
sum rules has indicated that both the form and the cut-off of its
form factor are not significantly different between a two-quark and
a four-quark $X(3872)$ meson \cite{navarra}.

The isospin- and spin-averaged cross section is then given by
\begin{eqnarray}  \label{cross}
\sigma_n=\frac{1}{64\pi sN_IN_S}\frac{p_f}{p_i}\int
d\Omega\overline{|\mathcal{M}_n|^2}F^4,
\end{eqnarray}
where $\overline{|\mathcal{M}_n|^2}$ denotes the squared amplitude
obtained from summing over the isospins and spins of both initial
and final particles, with $F$ denoting the appropriate form factors
at interaction vertices. The factors $N_I=(2I_1+1)(2I_2+1)$ and
$N_S=(2S_1+1)(2S_2+1)$ in the denominator are due to averaging over
the isospins $I_1$ and $I_2$ as well as the spins $S_1$ and $S_2$ of
initial particles. The three-momenta in the center of mass of
initial and final particles are denoted by $p_i$ and $p_f$,
respectively.

\begin{figure}[ht]
\includegraphics[width=2.8in,height=2.8in,angle=0]{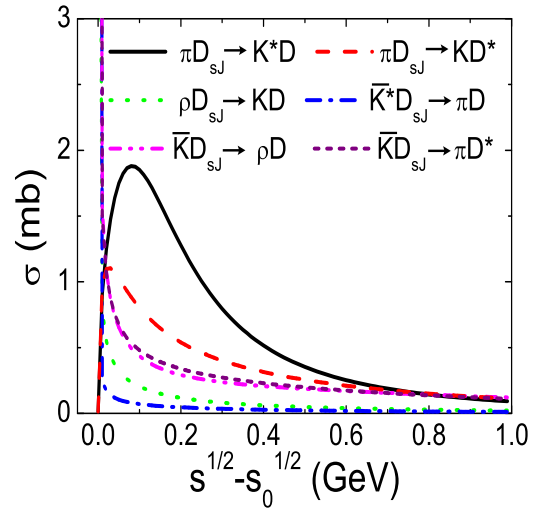} \vspace{0.5cm}
\caption{(color online) Cross sections for the absorption of a
four-quark $D_{sJ}$(2317) meson by $\protect\pi$, $\protect\rho$,
${\bar K}$, and ${\bar K}^*$ mesons via reaction $\protect\pi
D_{sJ}\to KD^*(K^*D)$, $\protect\rho D_{sJ}\to KD$, $\bar KD_{sJ}\to
\protect\rho D(\protect\pi D^*)$, and $\bar K^* D_{sJ}\to
\protect\pi D$.} \label{crosssection}
\end{figure}

In Fig.~\ref{crosssection}, we show the absorption cross sections of
the $D_{sJ}$(2317) meson by $\pi$, $\rho$, ${\bar K}$, and ${\bar
K}^*$ as functions of the total center-of-mass energy $s^{1/2}$
above the threshold energy $s_0^{1/2}$ of a reaction for the
scenario that it is a four-quark state. Aside near the threshold of
a reaction, where the cross section can be very large or small
depending on whether the reaction is exothermic or endothermic, most
cross sections are less than 1 mb except the reaction $\pi D_{sJ}\to
K^* D$, which has a peak value of about 2 mb. If the $D_{sJ}$(2317)
meson is a two-quark state, its absorption cross sections are about
a factor of nine larger than corresponding ones shown in
Fig.~\ref{crosssection} as the coupling constant $g_{D_{sJ}DK}$ is
about a factor of three larger for a two-quark $D_{sJ}$(2317) meson
than for a four-quark one.

The $D_{sJ}$(2317) meson can also be produced in the hadronic matter
by the inverse reactions $KD^*(DK^*)\to \pi D_{sJ}$, $KD\to\rho
D_{sJ}$, $\rho D(\pi D^*)\to \bar KD_{sJ}$, and $\pi D\to \bar K^*
D_{sJ}$, with cross sections related to those of absorption
reactions via the detailed balance relations.

\subsection{Thermally averaged $D_{sJ}$(2317) meson absorption cross sections}

In the kinetic model to be used in the next section for studying
$D_{sJ}$(2317) meson absorption and production in hadronic matter,
the thermally averaged cross sections are needed. In terms of the
thermal distribution functions $f_i(\mathbf{p})$ of $D_{sJ}$(2317)
mesons and other hadrons, the thermally averaged cross section
$\sigma _{ab\rightarrow cd}$ for the reaction $ab\to cd$ is given by
\cite{ko}
\begin{eqnarray}
\left\langle \sigma _{ab\rightarrow cd}v\right\rangle &=&\frac{\int
d^{3}\mathbf{p}_{a}d^{3}\mathbf{p}_{b}f_{a}(\mathbf{p}_{a})f_{b}(\mathbf{p}_{b})
\sigma _{ab\rightarrow cd}v_{ab}}{\int d^{3}\mathbf{p}_{a}d^{3}
\mathbf{p}_{b}f_{a}(\mathbf{p}_{a})f_{b}(\mathbf{p}_{b})}\nonumber\\
&=&[4\alpha^2_a K_2(\alpha_a) \alpha^2_b K_2(\alpha_b)]^{-1}  \nonumber \\
&\times&\int^\infty_{z_0} dz[z^2-(\alpha_a+\alpha_b)^2]
[z^2-(\alpha_a-\alpha_b)^2]\nonumber\\
&\times&K_1(z)\sigma(s=z^2 T^2),
\end{eqnarray}
with $\alpha_i=m_i/T$, $z_0=\mathrm{max}(\alpha_a+\alpha_b,\alpha_c
+\alpha_d)$, $K_1$ being the modified Bessel function, and $v_{ab}$
denoting the relative velocity of initial two interacting particles
$a$ and $b$, i.e.,
\begin{eqnarray}
v_{ab}=\frac{\sqrt{(p_a\cdot p_b)^2-m^2_am^2_b}}{E_aE_b}.
\end{eqnarray}

\begin{figure}[ht]
\includegraphics[width=2.8in,height=2.8in,angle=0]{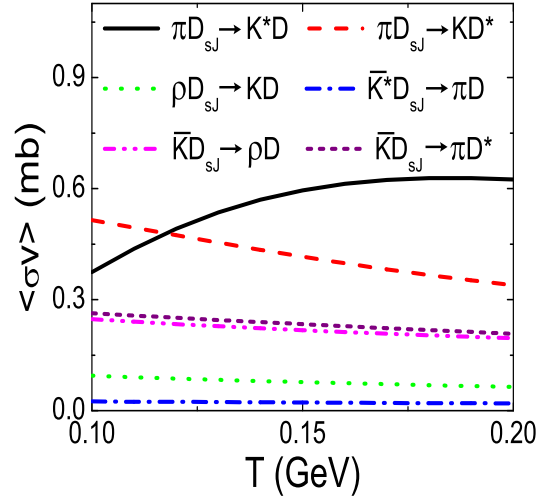} \vspace{0.5cm}
\caption{(color online) Thermally averaged cross sections for the
absorption of a four-quark $D_{sJ}$(2317) meson by $\protect\pi$,
$\protect\rho$, ${\bar K}$, and ${\bar K}^*$ mesons via reaction
$\protect\pi D_{sJ}\to KD^*(K^*D)$, $\protect\rho D_{sJ}\to KD$,
$\bar KD_{sJ}\to \protect\rho D(\protect\pi D^*)$, and $\bar K^*
D_{sJ}\to \protect\pi D$.} \label{sigmav}
\end{figure}

The thermally averaged absorption cross sections of the
$D_{sJ}$(2317) meson by $\pi$, $\rho$, ${\bar K}$, and ${\bar K}^*$
as functions of the temperature of the hadronic matter are shown in
Fig.~\ref{sigmav} for the case that $D_{sJ}$(2317) is a four-quark
meson. It is seen that the thermally averaged cross section of the
dominant reaction $\pi D_{sJ}\to K^* D$ has values less than 0.6 mb.
This value is again about a factor of nine larger if the
$D_{sJ}$(2317) is a two-quark meson.

\section{Time evolution of the $D_{sJ}$(2317) meson abundance in hadronic matter}
\label{kinetic}

\subsection{Rate equation for $D_{sJ}$(2317) meson production in heavy ion
collisions}

In terms of thermally averaged cross sections and the densities of
$\pi$, $\rho$, $K$, and $K^\ast$ mesons, the time evolution of the
$D_{sJ}$(2317) meson abundance in the hadronic matter is determined
by the kinetic equation
\begin{eqnarray}\label{rate}
\frac{dN_{D_{sJ}}(\tau)}{d\tau}&=&R_{QGP}(\tau)+\sum_{a,b,c}
\langle\sigma_{aD_{sJ}\to bc}v_{aD_{sJ}}\rangle n_a^{(0)}(\tau)\nonumber\\
&\times&\left[N_{D_{sJ}}^{(0)}(\tau)\frac{n_c(\tau)}{n_c^{(0)}(\tau)}
-N_{D_{sJ}}(\tau)\right],
\end{eqnarray}
where $n_a(\tau)$ and $n_c^{(0)}(\tau)$ are, respectively, the
equilibrium densities of light meson type $a$ and charmed meson type
$c$ in the hadronic matter at proper time $\tau$ when its
temperature is $T$ according to Eq.(\ref{temperature}), while
$N_{D_{sJ}}^{(0)}(\tau)$ is the equilibrium number of $D_{sJ}$(2317)
mesons given by Eq.(\ref{statistical}) using the temperature and
flow velocity at proper time $\tau$. Since hadronization of the
quark-gluon plasma takes a finite time of $\tau_H-\tau_C\simeq 2.5$
fm/$c$, $D_{sJ}$(2317) mesons are produced from the quark-gluon
plasma in the mixed phase, with a rate proportional to the volume of
the quark-gluon plasma. This is included in Eq.(\ref{rate}) through
the term $R_{QGP}(\tau)$. Since the fraction of the quark-gluon
plasma during the mixed phase decreases almost linearly with the
proper time, we can approximately write
\begin{equation}
R_{QGP}(\tau)=\cases{N_{D_{sJ}}^{(0)}/(\tau_H-\tau_C),&$\tau_C<\tau<\tau_H$;\cr
0,&Otherwise.\cr}
\end{equation}
In the above, $N_{D_{sJ}}^{(0)}$ is the total number of
$D_{sJ}$(2317) mesons produced from the quark-gluon plasma. In the
following calculations, it is obtained either from the coalescence
model by evaluating Eq.(\ref{coalmodel}) numerically using the
Monte-Carlo method or from the statistical model using
Eq.(\ref{statistical}). In writing Eq.(\ref{rate}), we have assumed
that the total number of charmed hadrons is conserved during the
evolution of the hadronic matter as charms are not likely to be
produced and destroyed in the hadronic matter because of their small
production and annihilation cross sections \cite{lin,liu1,liu2}.

\begin{figure}[ht]
\includegraphics[width=2.8in,height=2.8in,angle=0]{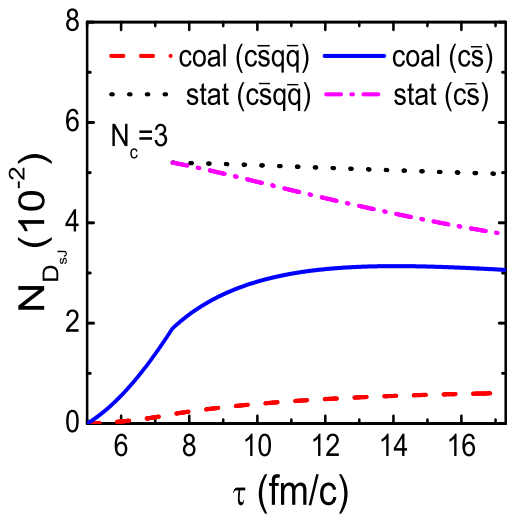}
\vspace{0.5cm} \caption{Time evolution of the $D_{sJ}$(2317) meson
abundance in central Au+Au collisions at $\protect\sqrt{s_{NN}}$=200
GeV for different initial numbers of $D_{sJ}$(2317) mesons produced
from the quark-gluon plasma.} \label{evolution}
\end{figure}

\subsection{$D_{sJ}$ meson yield in relativistic heavy ion collisions}

In Fig.\ref{evolution}, the abundance of $D_{sJ}$(2317) mesons in
central Au+Au collisions at $\sqrt{s_{NN}}=200$ GeV is shown as a
function of the proper time of the fire-cylinder. Since the initial
number of $D_{sJ}$(2317) mesons produced from the quark-gluon plasma
via quark coalescence is below the equilibrium number for both the
two-quark and four-quark $D_{sJ}$(2317) mesons, its number increases
during the hadronic evolution as shown by the dashed and solid
lines, respectively. The final number of $D_{sJ}$(2317) mesons is
about $3.0\times 10^{-2}$ if the $D_{sJ}$(2317) meson is a two-quark
state and is about $6.0\times 10^{-3}$ if it is a four-quark meson.
Although the ratio ($\sim 5$) between the final numbers for the two
and four-quark $D_{sJ}$(2317) mesons is smaller than that ($\sim
20$) for initially produced $D_{sJ}$(2317) mesons, it is still
appreciable. This result differs significantly from the predictions
of the statistical model. In this case, the $D_{sJ}$(2317) meson
number decreases slightly to $3.8\times 10^{-2}$ during the hadronic
evolution if it is a two-quark meson as shown by the dash-dotted
line, and it remains essentially unchanged during hadronic evolution
if it is a four-quark meson as shown by the dotted line. Since the
final yield of $D_{sJ}$(2317) mesons in the coalescence model is
much smaller for a four-quark state than for a two-quark state and
also that from the statistical model, studying $D_{sJ}$(2317) meson
production in relativistic heavy ion collisions thus provides the
possibility of understanding not only its production mechanism but
also its quark structure.

Above results are obtained by assuming that there are three charm
quarks in the quark-gluon plasma, based on the number of charm
quarks measured by the PHENIX collaboration in p+p collisions. If
this number is increased by a factor of two, which is closer to that
expected from the STAR experiment on d+Au collisions, the final
$D_{sJ}$(2317) meson numbers in both the coalescence and statistical
models are increased by about a similar factor. A similar reduction
factor is seen in the final $D_{sJ}$(2317) meson numbers in heavy
ion collisions if the total charm quark number is reduced by a
factor of two as given by the PYTHIA program for p+p collisions.

\section{summary}

\label{summary}

Using the quark coalescence model, we have predicted the yield of
$D_{sJ}$(2317) mesons in central Au+Au collisions at RHIC. Contrary
to the prediction of the statistical model, the initial number of
$D_{sJ}$(2317) meson produced at the end of the quark-gluon stage of
heavy ion collisions depends sensitively on whether it is a
two-quark or a four-quark meson, with the former giving an order of
magnitude larger number than the latter. To take into account the
effects of absorption and production during subsequent hadronic
evolution, we have used a hadronic model to evaluate the cross
sections for the absorption of the $D_{sJ}$(2317) meson by pion, rho
meson, anti-kaon, and vector anti-kaon in the tree-level Born
approximation. With empirical masses and coupling constants as well
as form factors from the QCD sum rules, we have found that all these
cross sections are small, except the reaction $\pi D_{sJ}\to K^*D$,
which has a peak cross section of about 2 mb, if the $D_{sJ}$(2317)
is a four-quark meson, but they are about ten times larger if it is
a conventional two-quark meson. Including these reactions in a
kinetic model based on a schematic hydrodynamic description of
relativistic heavy ion collisions, we have studied the time
evolution of the abundance of $D_{sJ}$(2317) mesons in these
collisions. Our results show that the large difference in the
initial numbers given by the quark coalescence model for the
two-quark and four-quark $D_{sJ}$(2317) mesons remains appreciable
at freeze out. On the other hand, the $D_{sJ}$(2317) number
determined from the statistical model is essentially unchanged if
the $D_{sJ}$(2317) is a four-quark meson and only change slightly if
it is a two-quark meson. Studying $D_{sJ}$(2317) production at RHIC
and also at the forthcoming LHC thus offers the possibility to
understand its quark structure and production mechanism. To achieve
this goal requires, however, accurate information on the number of
charm quarks produced during initial hard nucleon-nucleon
collisions, as increasing or decreasing the initial quark number by
a factor affects the final $D_{sJ}$(2317) meson number by a similar
factor. Also, it is important to know if charm quarks can be
produced from the quark-gluon plasma as this would affect the
initial $D_{sJ}$(2317) mesons produced from the quark-gluon plasma
as well.

\begin{acknowledgments}
We thank Yongseok Oh for helpful discussions. This paper was based
on work supported by the US National Science Foundation under Grant
No. PHY-0457265 and the Welch Foundation under Grant No. A-1358 (CMK
and WL), the NNSF of China under Grant Nos. 10575071 and 10675082,
MOE of China under project NCET-05-0392, Shanghai Rising-Star
Program under Grant No. 06QA14024, and the SRF for ROCS, SEM of
China (LWC), as well as CNPq and FAPESP (MN).
\end{acknowledgments}

\appendix

\section{Approximate evaluation of the coalescence integral}\label{analytic}

For a hypersurface of constant proper time and a distribution with
Bjorken correlation between $y$ and $\eta $, we can expand the
hyperbolic function in the coalescence integral to first order in
$y$ and $\eta$ if we consider $D_{s,J}$(2317) meson production at
midrapidity with $|y|<0.5$. In this case, the invariant phase-space
factor in Eq.(\ref{coal}) can be approximated by
\begin{equation}
p_{i}\cdot d\sigma _{i}\frac{d^{3}\mathbf{p}_{i}}{E_{i}}\simeq
d^{3}\mathbf{x}_{i}d^{3}\mathbf{p}_{i}.  \label{phase}
\end{equation}
Neglecting the transverse flow and treating quarks
nonrelativistically, we can then use the relation
\begin{equation}
\sum_{i=1}^n \frac{\mathbf{p}_i^2}{2\,m_i
T}=\frac{\mathbf{K}^2}{2M\,T}+
\sum_{i=1}^{n-1}\frac{\mathbf{k}_i^2}{2\mu_i T},
\end{equation}
where $M=\sum_{i=1}^n m_i$, to express the quark Boltzmann momentum
distribution functions in terms of the total ${\bf K}$ and relative
${\bf k}_i$ momenta. Using also the total and relative ${\bf y}_i$
coordinates of the quarks, we obtain the following expression for
the number of $D_{s,J}$(2317) mesons produced from quark
coalescence:
\begin{equation}
N_{D_{sJ}}=g_{D_{sJ}} \prod_{j=1}^{n}N_j \prod_{i=1}^{n-1}
\frac{\int d^3\mathbf{y}_i \,d^3\mathbf{k}_i\,
f_{D_{sJ}}^W(\mathbf{y}_i,\mathbf{k}_i) \,f_q(\mathbf{k}_{i})} {\int
d^3\mathbf{y}_i\, d^3\, \mathbf{k}_i\, f_q(\mathbf{k}_{i})}.
\label{coal2}
\end{equation}
In the above, we have made use of the fact the Wigner function of
$D_{sJ}$(2317) meson is factorable in both the relative coordinates
and momenta of its constituent quarks. Because of $-0.5\le y=\eta\le
0.5$, the momentum space integral in Eq.(\ref{coal2}) reduces to a
two-dimensional one, as the momentum integral in the $z$-direction
gives simply one.

If the $D_{sJ}$(2317) meson is a p-wave two-quark meson, evaluating
the integrals in Eq.(\ref{coal2}) with its Wigner function given by
Eq.(\ref{wigner1}) leads to the following number of $D_{sJ}$(2317)
mesons produced from quark coalescence:
\begin{eqnarray}\label{diquark}
N_{D_{sJ}}^{\rm two}&\simeq& \frac{1}{27}N_cN_{\bar s} \frac{(4\pi
)^{3/2}\mu T_C\sigma^5}{V_C(1+2\mu T_C\sigma^2)^2}\nonumber\\
&\approx& 1.9\times 10^{-2},
\end{eqnarray}

For a four-quark $D_{sJ}$(2317) meson with its Wigner function given
by Eq.(\ref{wigner2}), the number of $D_{sJ}$(2317) mesons produced
from quark coalescence is
\begin{eqnarray}\label{fourquark}
N_{D_{sJ} }^{\rm four}&\simeq& \frac{1}{1296}N_cN_{\bar
s}(N_uN_{\bar u}+N_dN_{\bar d})\nonumber\\
&\times&\prod_{i=1}^3\frac{(4\pi\sigma_i^2)^{3/2}}{V_C(1+2\mu
_{i}T_C\sigma_{i}^{2})}\approx 1.1\times 10^{-3}.
\end{eqnarray}

\section{The $D_{sJ}DK$ form factor}

\label{qcd}

In this appendix, we compute the $D_{sJ}DK$ form factor using the
QCD rum rules \cite{svz,rry}. In this approach, the short-range
perturbative QCD is extended by the Wilson's operator product
expansion (OPE) of the correlators, which results in a series in
powers of the squared momentum with Wilson coefficients. The
convergence at low momentum is improved by using a Borel transform.
The expansion involves universal quark and gluon condensates.
Equating the quark-based calculation of a given correlator to the
same correlator that is calculated using hadronic degrees of freedom
via a dispersion relation then provides the sum rules from which a
hadronic quantity can be estimated.

We shall use the three-point function to evaluate the $D_{sJ}DK$
form factor by following the procedure suggested in
Ref.~\cite{rhodd} and further extended in \cite{dsdpi}. This means
that we shall calculate the correlators for an off-shell $D$ meson
and then for an off-shell $K$ meson, requiring that the
corresponding extrapolations to the respective poles lead to the
same unique coupling constant.

The three-point function associated with a $D_{sJ}DK$ vertex with an
off-shell $D$ meson is given by
\begin{eqnarray}
\Gamma^{(D)}_{\mu}(p,p^\prime)&=&\int d^4x \, d^4y \, \langle
0|T\{j_{5\mu}(x) j_D(y)j^\dagger_{D_{sJ}}(0)\}|0\rangle  \nonumber \\
&\times&e^{ip^\prime.x} \, e^{i(p-p^\prime).y}\; ,  \label{cor}
\end{eqnarray}
where $j_{5\mu}=\bar{s}\gamma_\mu\gamma_5 q$, $j_D=i\bar{q}\gamma_5
c $ and $j_{D_{sJ}}=\bar{c}s$ are the interpolating fields for the
$K$, $D$, and $D_{sJ}$, respectively with $q$, $s$ and $c$ being the
light, strange and charm quark fields. Here we take the $D_{sJ}$ to
be a standard scalar quark-antiquark meson.

The phenomenological side of the vertex function, $\Gamma_{\mu}
(p,p^\prime)$, is obtained by the consideration of $K$ and $D$
states contribution to the matrix element in Eq.~(\ref{cor}):
\begin{eqnarray}
\Gamma_{\mu}^{(D){\rm phen}}(p,p^\prime)&=&{\frac{m_{D_{sJ}} m_{D}^2
F_K f_{D}
f_{D_{sJ}} }{m_c(p^2-m_{D_{sJ}}^2)({p^\prime}^2-m_K^2)}}  \nonumber \\
&\times&{\frac{g^{(D)}_{D_{sJ}DK}(q^2)}{(q^2-m_{D}^2)}}p^\prime_\mu
\nonumber \\
&+& \mbox{higher resonances}\; .  \label{phen}
\end{eqnarray}
In deriving Eq.~(\ref{phen}), we have made use of
\begin{equation}
\langle D_{sJ}(p)|K(p^\prime) D(q)\rangle = g^{(D)}_{D_{sJ}DK}(q^2),
\end{equation}
where $q=p^\prime-p$, and the decay constants $F_K$ and $f_{D}$ and
$f_{D_{sJ}}$ are defined by the matrix elements
\begin{equation}
\langle 0|j_{5\mu}|K(p^\prime)\rangle=ip^\prime_\mu F_K,  \label{fk}
\end{equation}
\begin{equation}
\langle 0|j_D|D(q)\rangle={\frac{m_{D}^2f_{D}}{m_c}} \; ,  \label{fd}
\end{equation}
and
\begin{equation}
\langle 0|j_{D_{sJ}}|D_{sJ}(p)\rangle={m_{D_{sJ}}f_{D_{sJ}}}.  \label{fdsj}
\end{equation}
The contribution of higher resonances and continuum in Eq.~(\ref{phen}) will
be taken into account as usual in the standard form of Ref.~\cite{io2},
through the continuum thresholds $s_0$ and $u_0$ for the $D_{sJ}$ and $K$
mesons, respectively.

The QCD side, or theoretical side, of the vertex function is
evaluated by performing Wilson's operator product expansion of the
operator in Eq.~(\ref{cor}). Expressing $\Gamma_{\mu}$ in terms of
the invariant amplitudes,
\begin{equation}
\Gamma_{\mu}(p,p^\prime)=F_1(p^2,{p^\prime}^2,q^2)p_\mu+F_2(p^2,{p^\prime}^2,q^2)
p^\prime_\mu,
\end{equation}
we can write a double dispersion relation for each one of the
invariant amplitudes, $F_i$, over the virtualities $p^2$ and
${p^\prime}^2$ holding $Q^2=-q^2$ fixed:
\begin{eqnarray}
&&F_i^{(D)}(p^2,{p^\prime}^2,Q^2)  \nonumber \\
&&=-{\frac{1}{4\pi^2}}\int_{m_c^2}^\infty ds \int_0^\infty du
{\frac{\rho_i(s,u,Q^2)}{(s-p^2)(u-{p^\prime}^2)}}\;,  \label{dis}
\end{eqnarray}
where $\rho_i(s,u,Q^2)$ equals the double discontinuity of the
amplitude $F_i(p^2,{p^\prime}^2,Q^2)$ on the cuts $m_c^2\leq
s\leq\infty$ and $0\leq u\leq\infty$, which can be evaluated using
Cutkosky's rules. Finally, in order to suppress the condensates of
higher dimension and at the same time reduce the influence of higher
resonances, we perform a double Borel transform in both variables
$P^2=-p^2\rightarrow M^2$ and ${P^\prime}^2=-{p^\prime}^2\rightarrow
{M^\prime}^2$. Equating the two representations described above, we
obtain the following sum rule in the structure $p^\prime_\mu $:
\begin{eqnarray}
& & \frac{m_{D_{sJ}} m^2_{D}}{m_c} F_{K} f_{D}f_{D_{sJ}} g^{(D)}_{D_{sJ}
DK}(Q^2) e^{ -m^2_{D_{sJ}}/M^{2} } e^{ - m^2_{K}/M^{^{\prime}2} }  \nonumber \\
&&= (Q^2 + m^2_{D}) \,\, \bigg[ m_c<\bar{s}s> e^{-m^2_c/M^{2}}  \nonumber \\
& & - \frac{1}{4 \pi^2} \int_{m^2_c}^{s_0} ds \int_0^{u_{max}} \, d u
\exp(-s/M^2) \exp(-u/M^{^{\prime}2})  \nonumber \\
&&\times f(s,t,u) \theta(u_0 -u) \bigg],  \label{dsoff}
\end{eqnarray}
where $t=-Q^2$ and
\begin{eqnarray}
f(s,t,u)&=&{\frac{3}{2[\lambda(s,u,t)]^{1/2}}}\left(m_c^2+2m_cm_s-s+\right.
\nonumber \\
&+&[(2m_c^2+2m_cm_s-s-t+u)(m_c^2(s-t+u)  \nonumber \\
&+&s(t+u-s)][\lambda(s,u,t)]^{-1},
\end{eqnarray}
with $\lambda(s,u,t)=s^2+u^2+t^2-2su-2st-2tu$, and
$u_{max}=s+t-m_c^2-{st/m_c^2}$.

We use the same parameters as in Ref.~\cite{wang2}:
$m_s=0.15\,\mathrm{GeV}$, $m_c=1.26\,\mathrm{GeV}$, $F_K=
0.16\,\mathrm{GeV}$, $m_{D}=1.865\,\mathrm{GeV}$,
$m_K=0.498\,\mathrm{GeV}$, $m_{D_{sJ}}=2.317\, \mathrm{GeV}$,
$f_{D}=0.23~\mathrm{GeV} $, $f_{D_{sJ}}=0.225\,\mathrm{GeV}$,
$\langle\overline{s}s \rangle\,=0.8\langle\overline{q}q\rangle$,
with $\langle\overline{q}q\rangle\,=\,-(0.245)^3\,\mathrm{GeV}^3$.
For the continuum thresholds we take
$s_0=(6.3\pm0.1)\,\mathrm{GeV}^2$ and $u_0=(m_K+\Delta u)^2$ with
$\Delta u=0.5\,\mathrm{GeV}$.

We also use the same Borel window as in Ref.~\cite{wang2}:
$10\,\mathrm{GeV}^2\leq M^2\leq 20\,\mathrm{GeV}^2$ and work at a
fixed ratio ${M^\prime}^2/M^2=0.64/m_{D_{sJ}}^2$. We find a good
Borel stability in this region of the Borel mass. Fixing
$M^2=15~\mathrm{GeV}^2$, we show in Fig.~6 by the filled circles the
momentum dependence of $g^{(D)}_{D_{sJ} DK}(Q^2)$.

\begin{figure}[tbp]
\label{formfactor}
\includegraphics[width=2.8in,height=2.8in]{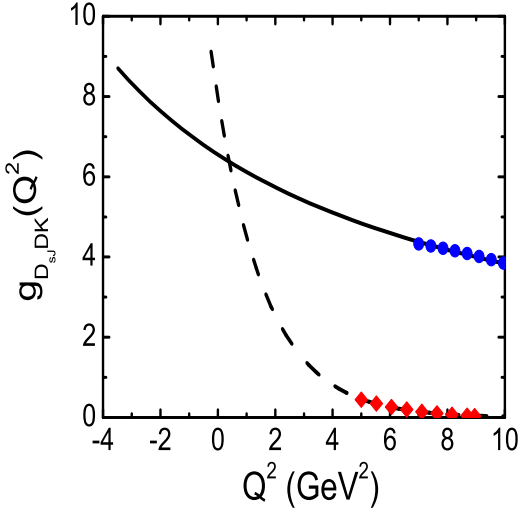}
\caption{Momentum dependence of the $D_{sJ} D K$ form factors. The
solid and dashed lines give the parametrization of the QCDSR results
for $g^{(D)}_{D_{sJ} DK}(Q^2)$ (circles) and $g^{(K)}_{D_{sJ}
DK}(Q^2)$ (squares), respectively.}
\end{figure}

Since the present approach can not be used at small values of $Q^2$,
extracting the $g_{D_{sJ} DK}$ coupling from the form factor
requires extrapolation of the curve to the mass of the off-shell
meson $D$. In order to do this, we fit the QCD sum-rule results with
an analytical expression. We have obtained a reasonable fit using a
monopole form:
\begin{equation}
g^{(D)}_{D_{sJ} DK}(Q^2)={\frac{92.4}{Q^2+14.1}}\;,  \label{gdoff}
\end{equation}
where the numbers are in units of $\mathrm{GeV}^2$. This fit is also
shown in Fig.~6 by the solid line. From Eq.(\ref{gdoff}) we get
$g_{D_{sJ} DK}=g^{(D)}_{D_{sJ} DK}(Q^2=-m_{D}^2)=8.7$.

To check the consistency of above fit, we also evaluate the form
factor at the same vertex, but for an off-shell kaon. In this case,
we have to evaluate the three-point function
\begin{eqnarray}
\Gamma^{(K)}_{\mu}(p,p^\prime)&=&\int d^4x \, d^4y \, \langle 0|T\{j_D(x)
j_{5\mu}(y)j^\dagger_{D_{sJ}}(0)\}|0\rangle  \nonumber \\
&\times& e^{ip^\prime.x} \, e^{i(p-p^\prime).y}\; .  \label{corpi}
\end{eqnarray}
Proceeding in a similar way, we obtain the following sum rule:
\begin{eqnarray}
&&\frac{m_{D_{sJ}} m^2_{D}}{m_c} F_{K} f_{D}f_{D_{sJ}}
g^{(K)}_{D_{sJ} DK}(Q^2) e^{ -m^2_{D_{sJ}}/M^{2} } e^{ -
m^2_{D}/M^{^{\prime}2} } \nonumber
\\
&&= - {\frac{Q^2 + m^2_{K}}{4 \pi^2}} \int_{m^2_c}^{s_0} ds
\int_{u_{min}}^{u_0} \, d u e^{-s/M^2} e^{-u/M^{^{\prime}2}}  \nonumber \\
&\times& g(s,t,u),  \label{koff}
\end{eqnarray}
where $u_{min}=m_c^2-{\frac{m_c^2t}{s- m_c^2}}$ and
\begin{eqnarray}
&&g(s,t,u)={\frac{3}{[\lambda(s,u,t)]^{3/2}}}[m_c^4(s-t+3u)  \nonumber \\
&&+u(m_cm_s(s+t-u)+s(-s+t+u)  \nonumber \\
&&+m_c^2(-2u(s-t+u)+m_cm_s(s-t+3u))].
\end{eqnarray}

Using now $u_0=(m_{D}+\Delta_u)^2$ with $\Delta_u=0.5\mathrm{GeV}$
and $M^{^{\prime}2}={\frac{m_D^2}{m_{D_{sJ}}^2}}M^2$, we find that
the results are also rather stable as a function of the Borel mass.
Fixing $M^2=15~\mathrm{GeV}^2$, we show in Fig.~6 by the squares the
QCD sum-rule results for $g^{(K)}_{D_{sJ} DK}(Q^2)$. A good fit of
these results can be obtained using an exponential form:
\begin{equation}
g^{(K)}_{D_{sJ} DK}(Q^2)=7.98 ~e^{-Q^2 / 1.75},  \label{gkoff}
\end{equation}
where 1.75 is in units of $\mathrm{GeV}^2$, as shown in Fig.~6 by
the dashed line. From Eq.(\ref{gkoff}) we get $g_{D_{sJ}
DK}=g^{(K)}_{D_{sJ} DK}(Q^2=-m_{K}^2)=9.1$, in excellent agreements
with both the result obtained from $g^{(D)}_{D_{sJ}
DK}(Q^2=-m_{D}^2)$ above and the result obtained in
Ref.~\cite{wang2} for this coupling constant: $g_{D_{sJ}
DK}=9.2\pm0.5$.

Considering the uncertainties in the continuum thresholds, and the
difference in the values of the coupling extracted when the $D$
meson or the kaon is off-shell, our result for the $D_{sJ} D K$
coupling constant is thus $g_{D_{sJ} DK}=8.9\pm0.9$.

>From the parameterizations in Eqs.(\ref{gdoff}) and (\ref{gkoff}),
we can also get information about the cut-off ($\Lambda$) in the
form factors. We see that the cut-off is much bigger when the $D$
meson is off-shell ($\Lambda\approx 3.7\,\mathrm{GeV}$) than when
the kaon is off-shell ($\Lambda\approx 1.3\,\mathrm{GeV}$), in
agreement with the results obtained in refs.~\cite{rhodd,dsdpi}.

\end{document}